# Chemical order in Ge-Ga-Sb-Se glasses

I. Pethes[a,*], R. Chahal[b,1], V. Nazabal[b], C. Prestipino[b], S. Michalik[c], J. Darpentigny[d], P. Jóvári[a]

[a]Wigner Research Centre for Physics, Hungarian Academy of Sciences, H-1525 Budapest, POB 49, Hungary

[b]Institut Sciences Chimiques de Rennes, UMR-CNRS 6226, Campus de Beaulieu, Université de Rennes 1, 35042 Rennes Cedex, France

[c]Diamond Light Source Harwell Science and Innovation Campus, Didcot, Oxfordshire, OX11 0DE, UK

[d]Laboratoire Léon Brillouin, CEA-Saclay 91191 Gif sur Yvette Cedex, France

Abstract

The short range order in $Ge_{30}Ga_5Sb_{10}Se_{55}$ and $Ge_{21}Ga_5Sb_{10}Se_{64}$ glasses was investigated by X-ray (XRD) and neutron diffraction (ND) as well as extended X-ray absorption fine structure (EXAFS) measurements at the Ge, Ga, Sb and Se K-edges. Large scale structural models were obtained by fitting simultaneously the experimental data sets by reverse Monte Carlo (RMC) simulation technique. It was found that Ge, Sb and Se atoms follow the Mott-rule and have 4, 3 and 2 nearest neighbors, respectively. The average coordination number of the Ga atoms was around 4. The structure of these glasses can be described by the chemically ordered network model: the Ge-Se, Ga-Se and Sb-Se bonds are the most prominent while Ge-Ge and Ge-Sb bonds are formed only in Se-poor compositions. Models generated by RMC contained some long distances (0.3-0.4 Å higher than the usual covalent bond lengths) between Ge-Se and/or Ge-Ge pairs. Dedicated simulation runs confirm the existence of these bonds.

---

[*] Corresponding author. E-mail address: pethes.ildiko@wigner.mta.hu
[1] present address: Department of Physics, NTNU, NO-7491 Trondheim, Norway

# 1. Introduction

Chalcogenide glasses are among the most frequently investigated materials because of their specific properties and promising possibilities of technological applications in many fields, such as infrared (IR) optics, photonics or medical utilizations. They have wide transparency window in the mid-infrared range, low phonon energy, high linear and nonlinear refractive indices and unique photosensitivity [1, 2]. Since their passive optical properties can be tuned by the chemical composition they are widely used in IR optics such as IR lenses, optical fibers and waveguides [1, 3]. Rare-earth (RE) doping makes them suitable for active optical usage, such as sensors, lasers or optical fiber amplifiers [4 - 8].

Selenide glasses are especially interesting due to their broad glass forming region, favorable thermal stability and wide transparency window [9 - 11]. Due to their lower toxicity antimony containing systems are more preferable for medical applications than compositions with arsenic. The addition of Ga enhances the rare-earth solubility of these glasses [12]. Therefore it is not surprising that vitreous Ge-Ga-Sb-Se alloys are extensively investigated [11, 13 - 16].

Chalcogenide glasses can be described as covalently bonded networks. The total coordination number ($N_i$) of the elements according to the Mott-rule [17] is equal to 8-$N$, where $N$ is the number of the valence electrons of the $i$th element. The validity of this rule is verified for groups 14, 15 and 16 of the periodic table (see e.g. [18 - 21]), but it was found that some group 13 elements deviate from the Mott-rule: Fourfold bonded Ga was reported in many publications [22 - 25], and total coordination number higher than three was found for In (e.g. in [26, 27]). Moreover, the total coordination number of the chalcogen atom can be higher than two in Ga and In containing glasses [24, 28].

There are several models about the bonding preferences of the constituent particles, such as the random network model (RNM) or the chemically ordered network model (CONM) [29]. While the $N_{ij}$ coordination numbers are determined only by the concentrations of the elements in the first model, there are frequent bonds and avoided bonds in the latter ($N_{ij}$ is the average number of $j$ type atoms around an $i$ type atom). According to the CONM model in the Ge-Ga-Sb-Se system the bonds between M and Se (M=Ge, Ga or Sb) are preferred, and the number of M-M and Se-Se bonds is minimized. The network is built from structural units, such as $GeSe_{4/2}$ or $SbSe_{3/2}$, which are connected by their corners and edges, or via Se-Se bridges in Se-rich glasses. In Se-poor systems where there are not enough Se atoms to satisfy the bonding requirements of the M atoms, M-M bonds are also expected to be present.

There are several studies about the structure of the ternary parent glasses (Ge-Ga-Se and Ge-Sb-Se) (see e.g. [21, 24] and references therein). Diffraction and extended X-ray absorption fine structure (EXAFS) measurements proved that the structure of Ge-Sb-Se glasses can be described by the CONM [21]. A similar study about the Ge-Ga-Se systems showed deviations from the Mott-rule not only for Ga, but even for Se, and violation of the chemical ordering was reported as well [24].

The presence of fourfold coordinated Ga-atoms was confirmed by Ga K-edge EXAFS measurements for the quaternary Ge-Ga-Sb-Se system [5, 12]. Raman experiments suggested $GeSe_{4/2}$, $GaSe_{4/2}$ and $SbSe_{3/2}$ structural units, as well as some violation of the chemical ordering in the form of Se-Se and M-M bonds [5, 14, 30]. But there is no comprehensive study on their atomic level structure yet, which would be helpful to better understand their properties.

In this paper X-ray diffraction measurements and EXAFS spectroscopy at Ge, Ga, Sb and Se K-edges of $Ge_{30}Ga_5Sb_{10}Se_{55}$ and $Ge_{21}Ga_5Sb_{10}Se_{64}$ glasses are presented. The $Ge_{21}Ga_5Sb_{10}Se_{64}$ glass was investigated additionally with neutron diffraction. The experimental data sets are fitted simultaneously in the framework of the reverse Monte Carlo (RMC) simulation method. Large scale three dimensional atomic models are obtained and the short range order parameters are determined.

## 2. Experimental

Sample preparation and density measurement are described in [21]. Densities of the glasses are shown in Table 1.

*2.1. X-ray diffraction measurements*

X-ray diffraction data were measured in transmission geometry at the Joint Engineering, Environmental and Processing (I12-JEEP) beamline [31] at Diamond Light Source Ltd (UK). The beam size was $0.3 \times 0.3$ mm$^2$. The wavelength of the monochromatic beam, sample-to-detector distance, position of beam center on the 2D detector and the tilt of the detector with respect to the direct beam were determined by measuring a $CeO_2$ standard [32]. The wavelength of incident photons was 0.14831 Å (83.595 keV) while the sample-to-detector distance was 336 mm.

Powder samples were filled into thin walled (0.01 mm) quartz capillaries with diameter of 1 mm. Scattered intensities were measured by a large area 2D detector (Thales Pixium RF4343). A single scattering pattern (image) was measured for 20 s. For both samples 45 images were summed up to obtain good statistics at higher scattering angles. Scattering of an empty capillary was also measured and subtracted directly from sample images. The DAWN software [33] was used to integrated 2D patterns into $Q$-space. The $S(Q)$ total structure factors were obtained after correcting integrated intensities for self-absorption, Compton scattering, fluorescence and multiple scattering by the PDFGetX2 program [34].

*2.2. Neutron diffraction measurement*

The neutron diffraction experiment was carried out at the 7C2 liquid and amorphous diffractometer of LLB (Saclay, France) [35]. Detector position and wavelength of monochromatic neutrons were determined by measuring a Ni standard. The wavelength of incident radiation was 0.723 Å. Powdered sample was placed into a thin walled cylindrical V sample holder of 5 mm diameter. Scattered intensities were recorded by the new 3He detector. Raw data were corrected for detector efficiency and background scattering.

*2.3. EXAFS measurements*

Ge, Ga, Se and Sb K-edge EXAFS data were measured in transmission mode at the GILDA-BM08 beamline of the ESRF (Grenoble, France). Monochromatic beam was obtained by a Si (311) fixed-exit double crystal monochromator. Higher harmonics were removed by two Pd coated mirrors set at an incidence angle of 3.6 mrad. Ionization chambers filled with Ar gas at different pressures were used to measure intensities of the incident and transmitted beam. Finely ground samples were mixed with cellulose powder and pressed into pellets. For each tablet the absorption $\mu t$ was about 1.5 above the selected absorption edge.

$\chi(k)$ curves were obtained from raw absorption spectra by the VIPER program [36]. $k^3$-weighted $\chi(k)$ signals were Fourier-transformed into *r*-space using a Kaiser-Bessel window ($\alpha$=1.5). The *k*-range of transformation was 1.85 Å$^{-1}$ -16 Å$^{-1}$ for the Ge and Se edges, 1.85 Å$^{-1}$ -12 Å$^{-1}$ for the Ga edge and 1.85 Å$^{-1}$ -14 Å$^{-1}$ for the Sb edge. The *r*-space data were then back-transformed using a rectangular window (over the *r*-space range 1.1 Å-2.8 Å for Ge, Ga and Se edges and 1.4 Å-3.1 Å for Sb edge data). Experimental total structure factors ($S(Q)$ functions) and $k^3$ weighted, filtered EXAFS signals ($\chi(k)$) are presented in Figures 1 and 2.

## 3. Reverse Monte Carlo simulations

The Reverse Monte Carlo simulation method [37] is a powerful tool to get large three dimensional atomic configurations which are consistent with measurements [38, 39]. Mostly structure factors from neutron and X-ray diffraction measurements and EXAFS data sets are fitted but in principle any experimental data can be used if they can be interpreted starting from the atomic coordinates. The incontrovertible gain of this method is that the experimental data sets are fitted simultaneously and the obtained atomic configurations are compatible with all of the measurements within their experimental errors. Moreover, additional physical and chemical information, such as density or preferred coordination numbers, can be taken into account as well.

In the RMC modeling partial pair correlation functions ($g_{ij}(r)$), $S(Q)$ and $\chi(k)$ curves are calculated from the atomic coordinates according to equations (1-3):

$$S(Q) = \sum_{i \leq j} w_{ij}^{X,N}(Q) S_{ij}(Q) \tag{1}$$

$$S_{ij}(Q) - 1 = \frac{4\pi\rho_0}{Q} \int_0^\infty r(g_{ij}(r) - 1)\sin(Qr)dr \tag{2}$$

$$\chi_i(k) = \sum_j 4\pi\rho_0 c_j \int_0^R r^2 \gamma_{ij}(k,r) g_{ij}(r) dr . \tag{3}$$

Here $S_{ij}(Q)$ are the partial structure factors, $w_{ij}^{X,N}(Q)$ are the weighting factors, which can be calculated from the concentrations and the coherent neutron scattering lengths for neutron and the atomic form factors for X-ray diffraction, respectively. $Q$ is the amplitude of the scattering vector, $\rho_0$ is the average number density, $i$ and $j$ denotes the considered atomic type. In the equation of the EXAFS signal $i$ refers to the absorber atom, $k$ is the wavenumber of the photoelectron, $\gamma_{ij}(k,r)$ is the photoelectron backscattering matrix, which gives the $k$-space contribution of a $j$-type backscatterer at distance $r$ from the absorber atom. Elements of $\gamma_{ij}(k,r)$ matrix were calculated for each $i$-$j$ pair by the FEFF8.4 program [40].

During the simulation particles are moved around randomly to fit experimental ($S_{exp}(Q)$, $\chi_{exp}(k)$) curves. The quality of the fits can be compared via their $R$-factors, which is defined for a given data set as:

$$R = \frac{\sqrt{\sum_i (S_{mod}(Q_i) - S_{exp}(Q_i))^2}}{\sqrt{\sum_i S_{exp}^2(Q_i)}}, \tag{4}$$

where $Q_i$ denotes the experimental points, $S_{mod}(Q)$ and $\chi_{mod}(k)$ are the model curves.

From the obtained particle configurations short range order parameters (nearest neighbor distances, average coordination numbers, bond angle distributions etc.) can be calculated.

In this study the RMC++ code [41] was used. The simulation boxes contained 10000 atoms for test runs and 30000 atoms for the final runs used for detailed analysis. The densities of the glasses are shown in Table 1. Initial configurations were obtained by placing the atoms into the simulation box randomly and moving them around to satisfy the minimum interatomic distance (cutoff distance) requirements (Table 2). Starting values of the cut off distances were usually around 90% of the sum of corresponding atomic radii [42]. The 10 partial pair correlation functions of a four component system can only be separated by fitting multiple datasets and using physical constraints, which can be used to reduce the uncertainty of structural parameters. One possibility is to forbid one or more bond types by using cutoff-distances higher than the typical (expected) bond distances. During the test runs several models were investigated with different combinations of allowed and forbidden bond types. Table 2 shows the cutoff-distances of all atomic pairs for both allowed and forbidden bonds. According to the CONM the Ge-Se, Ga-Se and Sb-Se bonds are preferred, thus these bond types were always allowed. In the final runs the first coordination shell of the $Ge_{21}Ga_5Sb_{10}Se_{64}$ sample consisted of Ge-Ge, Ge-Se, Sb-Se and Ga-Se contributions while Ge-Sb bonds were also present in $Ge_{30}Ga_5Sb_{10}Se_{55}$.

The $\sigma$ parameters used to calculate the RMC cost function [37] were reduced in three steps to the final values of $5\text{-}10\times10^{-4}$ for the diffraction data sets and $1\text{-}2\times10^{-5}$ for the EXAFS data sets. The number of accepted moves was typically around $1\text{-}2\times 10^7$.

The average coordination numbers ($N_{ij}$) are calculated from the partial pair correlation functions integrating them up to the minimum of the curves between the first and second coordination shells. The uncertainty of the $N_{ij}$ values is estimated by dedicated simulation runs, where the investigated average coordination number was changed systematically by applying constraints for them while the $R$-factors were monitored.

Random atomic moves leading to unrealistically high (e. g. 6 or more for Ge) and low (e.g. 0, 1 or 2 for Ge) coordination numbers were always rejected without calculating the $R$-factor.

4. Results and discussion

If the Mott-rule is satisfied then glassy $Ge_{21}Ga_5Sb_{10}Se_{64}$ is slightly Se-deficient, even if the Ga atoms are threefold coordinated. According to the CONM, mostly Ge-Se, Ga-Se and Sb-Se bonds are expected, Se-Se and M-M bonds would be present only in small amounts. Several test runs were made to determine the necessary bond types. Since this composition is nearly stoichiometric, a model in which only Ge-Se, Ga-Se and Sb-Se bonds were allowed, was also tested. It was found, that while the Se-Se bonds can be eliminated, at least one of the M-M type bonds is required to get proper fits. As the average of the $R$-factors of the fits was the best when the Ge-Ge bonds were allowed, the necessary bond type is most probably the Ge-Ge bond. The M-M type bond could be Ge-Ga as well, which was preferred by the Ga EXAFS data, but the uncertainty here is rather high due to the similar size and scattering properties of Ge and Ga and the low concentration of the latter. The quality of the fits obtained by applying the model in which only Ge-Ge, Ge-Se, Ga-Se and Sb-Se bonds were allowed are shown in Fig. 1.

The Se-deficiency of the $Ge_{30}Ga_5Sb_{10}Se_{55}$ glass is stronger, thus M-M type bonds need to be present beside the Ge-Se, Ga-Se and Sb-Se bonds to satisfy the bonding requirements, while the number of Se-Se bonds is expected to be negligible in this glass. Test runs showed that Ga-Ga, Ga-Sb, Sb-Sb and Se-Se bonds can be completely eliminated in this glass as well. The strong Se-deficiency required the presence of Ge-Sb pairs, the quality of the fits deteriorates without this bond: the *R*-factors increased an average 37 % in the model in which the Ge-Sb bonds were forbidden compared to the one with Ge-Sb bonds. The environment of Sb atoms consists of about one Ge and on average two Se atoms. The Ge-Ge pairs are also necessary to get adequate fits of the data sets, their elimination results in an average 20% worsening of the fits, especially the quality of the Se EXAFS fit decreased strongly (58%). Four of the five data sets can be properly fitted without Ge-Ga bonds, only the Ga EXAFS fit is sensitive to the elimination of Ge-Ga pairs. The quality of the fits obtained by applying the model, in which Ge-Ge, Ge-Sb, Ge-Se, Ga-Se and Sb-Se bonds were allowed are shown in Fig. 2.

The partial pair correlation functions are presented in Figs. 3 and 4. The $g_{ij}(r)$ curves have sharp peaks in the first coordination shell in the region 2.2 Å ≤ $r$ ≤ 2.8 Å. The nearest neighbor distances (locations of the first peaks) are collected in Table 3.

The Ge-Se (2.36-2.37 Å) and Ga-Se (2.39-2.41 Å) bond distances are similar to those found earlier in Ge-Ga-Se glasses (e.g. 2.36 Å / 2.38 Å [24] and 2.37 Å / 2.41 Å [43]). Neutron diffraction with isotopic substitution study of glassy $GeSe_2$ also gave 2.36 Å for the mean Ge-Se distance [44]. The Ge-Ge bond length (2.45-2.46 Å) is somewhat longer than what was found in Ge-As-Se glasses by X-ray diffraction and EXAFS measurements (2.36-2.42 Å [45]), in Ge-Sb-Se glasses (2.42-2.43 Å [21]) or in vitreous $GeSe_2$ glass (2.42 Å [44]). The higher value of the Ge-Ge bond length may be caused by the admixture of Ge-Ga distances that are supposed to be somewhat longer (≈2.48 Å [24]). First-principles molecular dynamic simulations resulted 2.47 Å Ge-Ge bond distance in $Ge_2Se_3$ glass [46].

The Sb-Se bond distance (2.58-2.59 Å) is somewhat lower than those reported earlier in Ge-Sb-Se films by EXAFS measurements (2.62 Å [47]) or in Sb-Se glasses determined by pulsed neutron diffraction experiments (2.64 Å, [48]), but similar to the value obtained in Ge-Sb-Se glasses by ND, XRD and EXAFS measurements and RMC simulations (2.58-2.60 Å [21]).

The Ge-Sb bond distance (2.65 Å) agrees well with the values found in Ge-Sb-S glasses by high energy synchrotron X-ray diffraction measurements (2.65 Å [49]) or in $Ge_{20}Sb_{20}Se_{60}$ (2.64 Å [21]).

The $N_{ij}$ average coordination numbers are listed in Table 4. The $N_i = \sum_j N_{ij}$ total coordination numbers of Ge, Sb and Se in the $Ge_{21}Ga_5Sb_{10}Se_{64}$ glass are 3.9, 2.9 and 1.82, respectively. The same values for the $Ge_{30}Ga_5Sb_{10}Se_{55}$ sample are 3.5, 3.1 and 1.98. The values for Sb and Se, which were obtained by simulation *without any coordination constraints*, are very close to the values predicted by the Mott-rule. The total coordination number of gallium is 4.1 and 4.0 in $Ge_{21}Ga_5Sb_{10}Se_{64}$ and $Ge_{30}Ga_5Sb_{10}Se_{55}$, respectively, which agrees well with the previous results in Ge-Ga-Se [24] and Ge-Ga-Sb-Se [5, 12] glasses.

The second coordination shell is located in the 3.3 Å $\leq r \leq$ 4.4 Å region. Some small peaks or shoulders can be observed in certain partials between the first and the second shell, in the 2.7 Å $< r <$ 3 Å region. Several test runs were carried out to judge the significance of these peaks. In these simulations zero coordination constraints were used in the 2.7 Å $< r <$ 3 Å region. It was found that the complete elimination of the 'long bonds' causes a significant increase of the $R$-factors (15-27 % on average), particularly of the ND fit of the $Ge_{21}Ga_5Sb_{10}Se_{64}$ glass (82 %) and the XRD fit of the $Ge_{30}Ga_5Sb_{10}Se_{55}$ glass (66%).

The role of the different partials was tested by allowing them one by one. It was found that these longer distances originate from Ge and/or Se related units. A small improvement of the quality of the fits can be observed when only Ge-Ge pairs are allowed in the above region. It is known that edge sharing $GeSe_4$ tetrahedra can be found in crystalline $GeSe_2$ [50]. However, the Ge-Ge distance in these tetrahedra is 3.049 Å while the small peak of $g_{GeGe}(r)$ appears at 2.8 Å in glassy $Ge_{21}Ga_5Sb_{10}Se_{64}$. In addition, improvement of fit qualities is more pronounced if Ge-Se bonding is allowed in this region. The fits of the data sets and partial pair correlation functions shown in Figs. 1 - 4 were obtained in simulations in which only the long Ge-Se pairs were allowed.

It often happens that in RMC-generated models of covalent glasses partial pair correlation functions have small non-zero values after the well-defined nearest neighbor peak. These long bonds are in most cases artificial and can be easily removed by using zero constraints after the first peak. This is not the case for glassy $Ge_{21}Ga_5Sb_{10}Se_{64}$ and $Ge_{30}Ga_5Sb_{10}Se_{55}$. Comparison of the Ge-Se partial pair correlation function of $Ge_{21}Ga_5Sb_{10}Se_{64}$ glass with those of $Ge_{27.5}As_{10}Se_{62.5}$ [45], $Ge_{20}Sb_{15}Se_{65}$ [21] and $Ge_{20}Ga_{10}Se_{70}$ glasses [24] obtained with the same methodology (Figure 5) also shows that the secondary peak between 2.7 Å and 3 Å is significant. The ratio of coordination numbers corresponding to 'normal' and long bonds is 0.2, 0.12, 0.055, 0.04 and 0.05 for $Ge_{30}Ga_5Sb_{10}Se_{55}$, $Ge_{21}Ga_5Sb_{10}Se_{64}$, $Ge_{27.5}As_{10}Se_{62.5}$, $Ge_{20}Sb_{15}Se_{65}$ and $Ge_{20}Ga_{10}Se_{70}$, respectively. While 0.12 and especially 0.2 is definitely larger than the relative error of $N_{GeSe}$ of the Ge-Ga-Sb-Se glasses the last three values are smaller than the relative error of the corresponding coordination numbers. This fact and the above discussed increase of $R$-factors both suggest that the presence of 'long bonds' is not a simulation artifact but results from the data used to obtain the models of Ge-Ga-Sb-Se glasses. We note here that long bonds were also found around Ge in $GeTe_4$-AgI glasses [51].

The Ge-Se coordination numbers were calculated up to $r_{max}$ = 2.7 Å ('short bonds') as well as between 2.7 Å < $r$ < 3 Å ('long bonds') (see Table 4). Taking into account the 'long bonds' for the calculation of the $N_{Ge}$ total coordination number too, $N_{Ge}$ is closer to 4 in the $Ge_{30}Ga_5Sb_{10}Se_{55}$ glass. The $N_{Ge}$ of the $Ge_{21}Ga_5Sb_{10}Se_{64}$ glass and the $N_{Se}$ values of both Ge-Ga-Sb-Se glasses together with the 'long bonds' are very close to the Mott-values.

## 5. Conclusions

The structure of $Ge_{30}Ga_5Sb_{10}Se_{55}$ and $Ge_{21}Ga_5Sb_{10}Se_{64}$ glasses was studied by X-ray diffraction, neutron diffraction and extended X-ray absorption fine structure (EXAFS) measurements at the Ge, Ga, Sb and Se K-edge. Reverse Monte Carlo simulation technique was used to fit the experimental data sets simultaneously. Short range order parameters were analyzed by using the obtained three dimensional structural models. It was found that the Ge, Sb and Se elements satisfy the Mott-rule and their coordination numbers are 4, 3 and 2, respectively. The Ga atoms have 4 nearest neighbors as was found earlier in Ga-containing ternary chalcogenide glasses. The structure of the examined glasses can be described by the chemically ordered network model. Ge-Se, Ga-Se and Sb-Se bonds are the most frequent. Chalcogen deficiency leads to the presence of M-M type bonds. The less Se-deficient sample can be modeled by allowing only one M-M type bond (Ge-Ge bonds were actually used but contribution of Ge-Ga pairs can not be excluded here). In $Ge_{30}Ga_5Sb_{10}Se_{55}$ Ge-Sb bonding is also necessary to obtain proper fits of the data sets. The other three M-M type bonds (Ga-Ga, Ga-Sb and Sb-Sb) can be completely eliminated. These observations are consistent with the previous results reported about the ternary Ge-Ga-Se [24] and Ge-Sb-Se [21] glasses. Some interatomic distances longer than the usual bond lengths were observed between 2.7 Å and 3.0 Å.

## Acknowledgments

I. P. and P. J. were supported by NKFIH (National Research, Development and Innovation Office) Grant No. SNN 116198. The neutron diffraction experiment was carried out at the ORPHÉE reactor, Laboratoire Léon Brillouin, CEA-Saclay, France. The authors thank to ESRF, (Grenoble, France) for the access to GILDA-BM08 beamline, where the EXAFS measurements were carried out. The authors are indebted to Angela Trapananti (University of Perugia) for her invaluable help during the EXAFS experiment. We also thank Diamond Light Source for access to beamline I12-JEEP that contributed to the results presented here.

Tables

Table 1 Mass and number densities of the investigated glass compositions.

| Composition | Mass density (g/cm$^3$) | Number density (atoms/Å$^3$) |
|---|---|---|
| $Ge_{30}Ga_5Sb_{10}Se_{55}$ | 4.838 ± 0.001 | 0.03602 |
| $Ge_{21}Ga_5Sb_{10}Se_{64}$ | 4.635 ± 0.001 | 0.03427 |

Table 2 Minimum interatomic distances (cut-offs) applied in the simulation runs (in Å).

| pair | Ge-Ge | Ge-Ga | Ge-Sb | Ge-Se | Ga-Ga | Ga-Sb | Ga-Se | Sb-Sb | Sb-Se | Se-Se |
|---|---|---|---|---|---|---|---|---|---|---|
| allowed | 2.25 | 2.3 | 2.45 | 2.15 | 2.3 | 2.45 | 2.15 | 2.65 | 2.4 | 2.1 |
| forbidden | 3.15 | 3.15 | 3.15 | - | 3.15 | 3.15 | - | 3.35 | - | 2.9 |

Table 3 Nearest neighbor distances (in Å) in the investigated glasses. The uncertainty of distances is usually ± 0.02-0.03 Å

| Pair | $Ge_{21}Ga_5Sb_{10}Se_{64}$ | $Ge_{30}Ga_5Sb_{10}Se_{55}$ |
|---|---|---|
| Ge-Ge | 2.45 | 2.46 |
| Ge-Sb | - | 2.65 |
| Ge-Se | 2.36 | 2.37 |
| Ga-Se | 2.39 | 2.41 |
| Sb-Se | 2.59 | 2.58 |

Table 4 Coordination numbers of the investigated glasses obtained by RMC simulation. In these simulation runs coordination constraints were not used. (For the Ge-Se partials the coordination numbers are calculated without and with 'long bonds' as well.)

|  | $Ge_{21}Ga_5Sb_{10}Se_{64}$ | $Ge_{30}Ga_5Sb_{10}Se_{55}$ |
|---|---|---|
| $N_{GeGe}$ | 0.7 (±0.5) | 0.9 (±0.5) |
| $N_{GeGa}$ | 0 | 0.0 |
| $N_{GeSb}$ | 0 | 0.35 (-0.1+0.3) |
| $N_{GeSe}$ | 3.2 / 3.6 (-0.1+0.3) | 2.25 / 2.7 (±0.2) |
| $N_{GaGe}$ | 0 | 0 |
| $N_{GaGa}$ | 0 | 0 |
| $N_{GaSb}$ | 0 | 0 |
| $N_{GaSe}$ | 4.1 (-0.5+1) | 4.0 (±1.0) |
| $N_{SbGe}$ | 0 | 1.05 (-0.3+0.9) |
| $N_{SbGa}$ | 0 | 0 |
| $N_{SbSb}$ | 0 | 0 |
| $N_{SbSe}$ | 2.9 (-0.5 +0.2) | 2.05 (-0.5+0.3) |
| $N_{SeGe}$ | 1.05 / 1.17 (±0.1) | 1.25/1.45 (±0.1) |
| $N_{SeGa}$ | 0.32 (±0.05) | 0.36 (±0.1) |
| $N_{SeSb}$ | 0.45 (-0.1 +0.05) | 0.37 (±0.1) |
| $N_{SeSe}$ | 0 | 0 |
| $N_{Ge}$ | 3.9 / 4.25 | 3.5/3.95 |
| $N_{Ga}$ | 4.1 | 4.0 |
| $N_{Sb}$ | 2.9 | 3.05 |
| $N_{Se}$ | 1.82 / 1.94 | 1.98/2.17 |

Figures

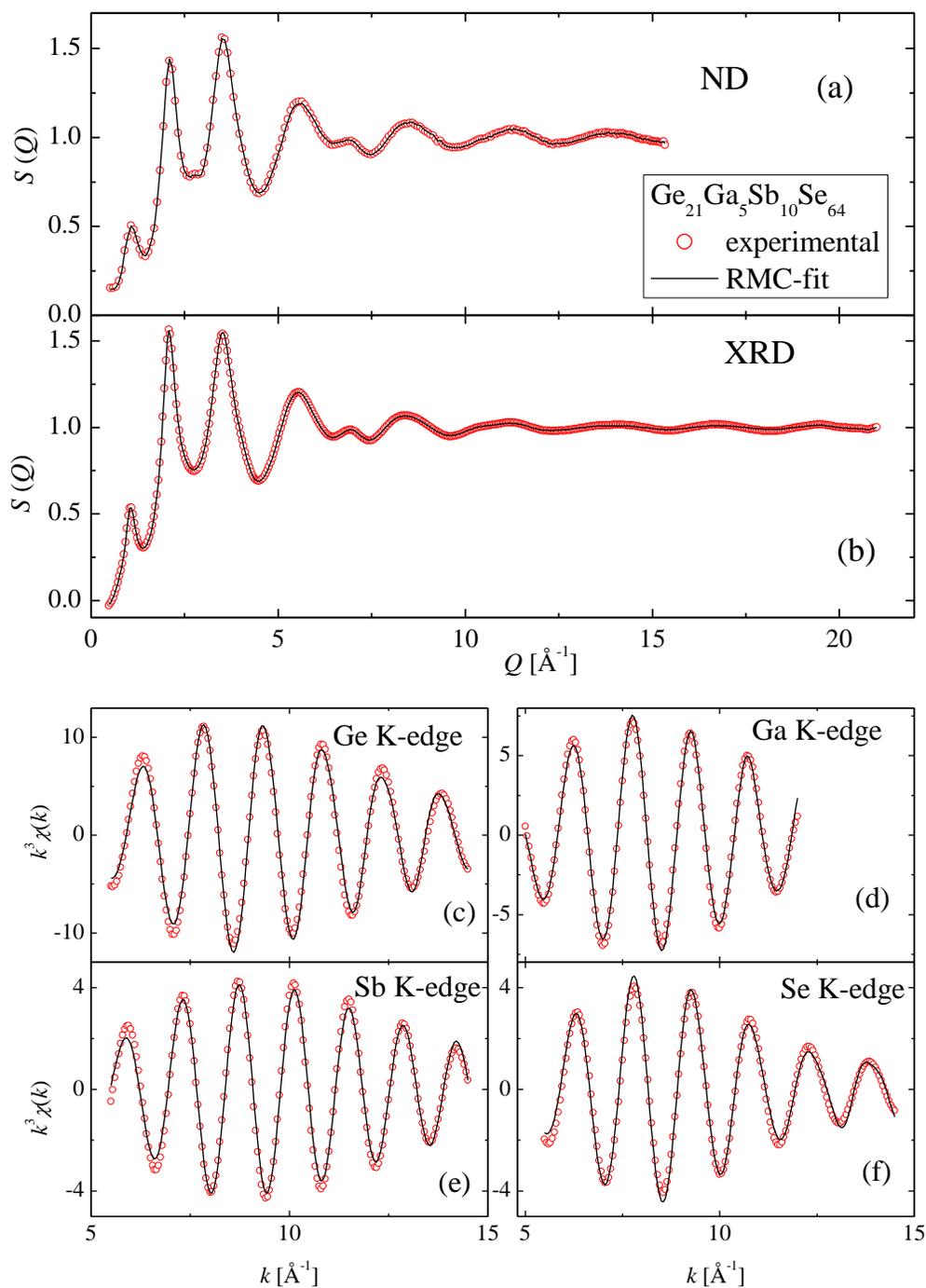

Figure 1 Neutron and X-ray diffraction structure function and $k^3$ weighted, filtered Ge, Ga, Sb and Se K-edge EXAFS spectra (open symbols) and fits (lines) of the $Ge_{21}Ga_5Sb_{10}Se_{64}$ sample.

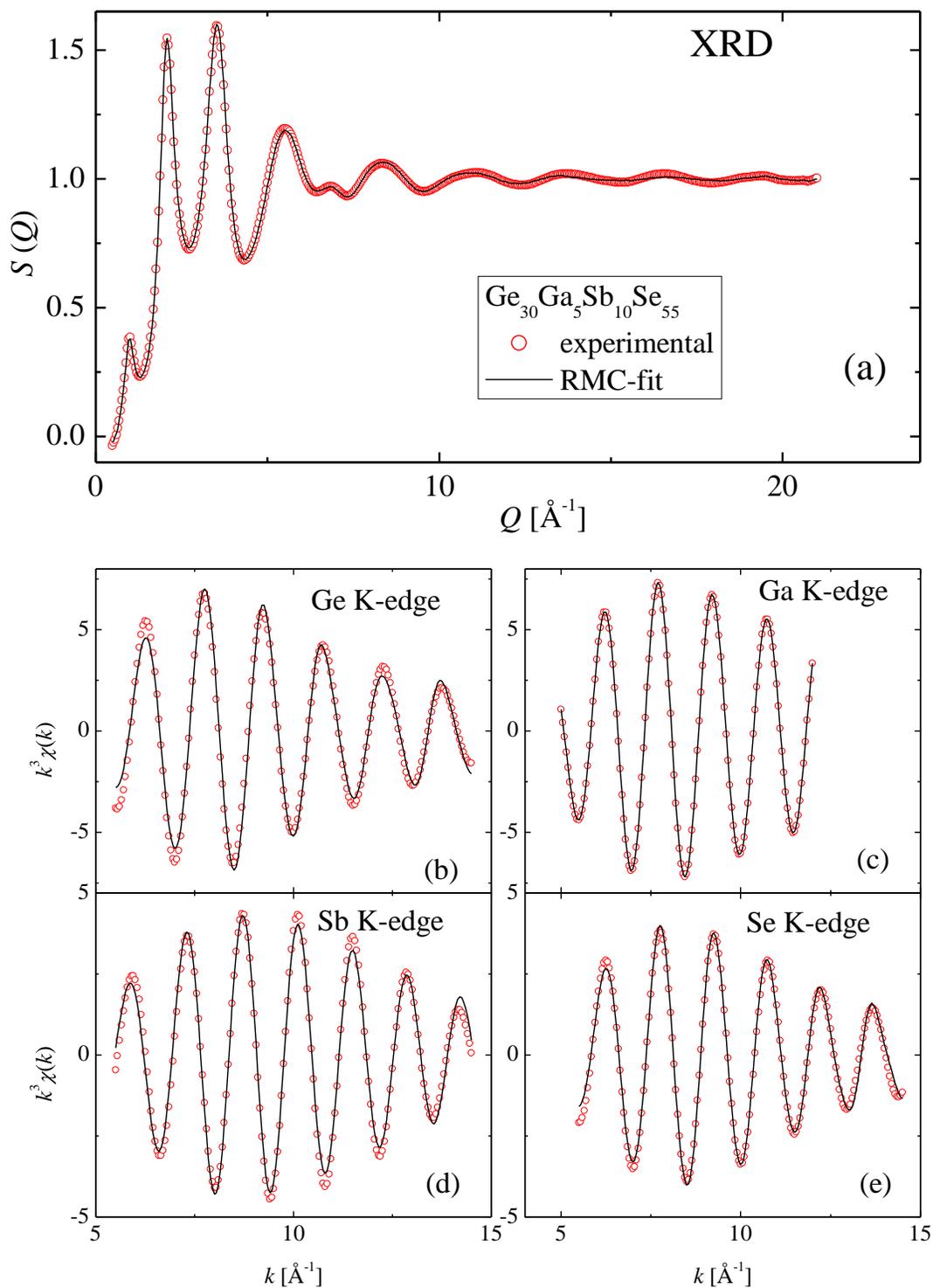

Figure 2 X-ray diffraction structure function and $k^3$ weighted, filtered Ge, Ga, Sb and Se K-edge EXAFS spectra (open symbols) and fits (lines) of the $Ge_{30}Ga_5Sb_{10}Se_{55}$ sample.

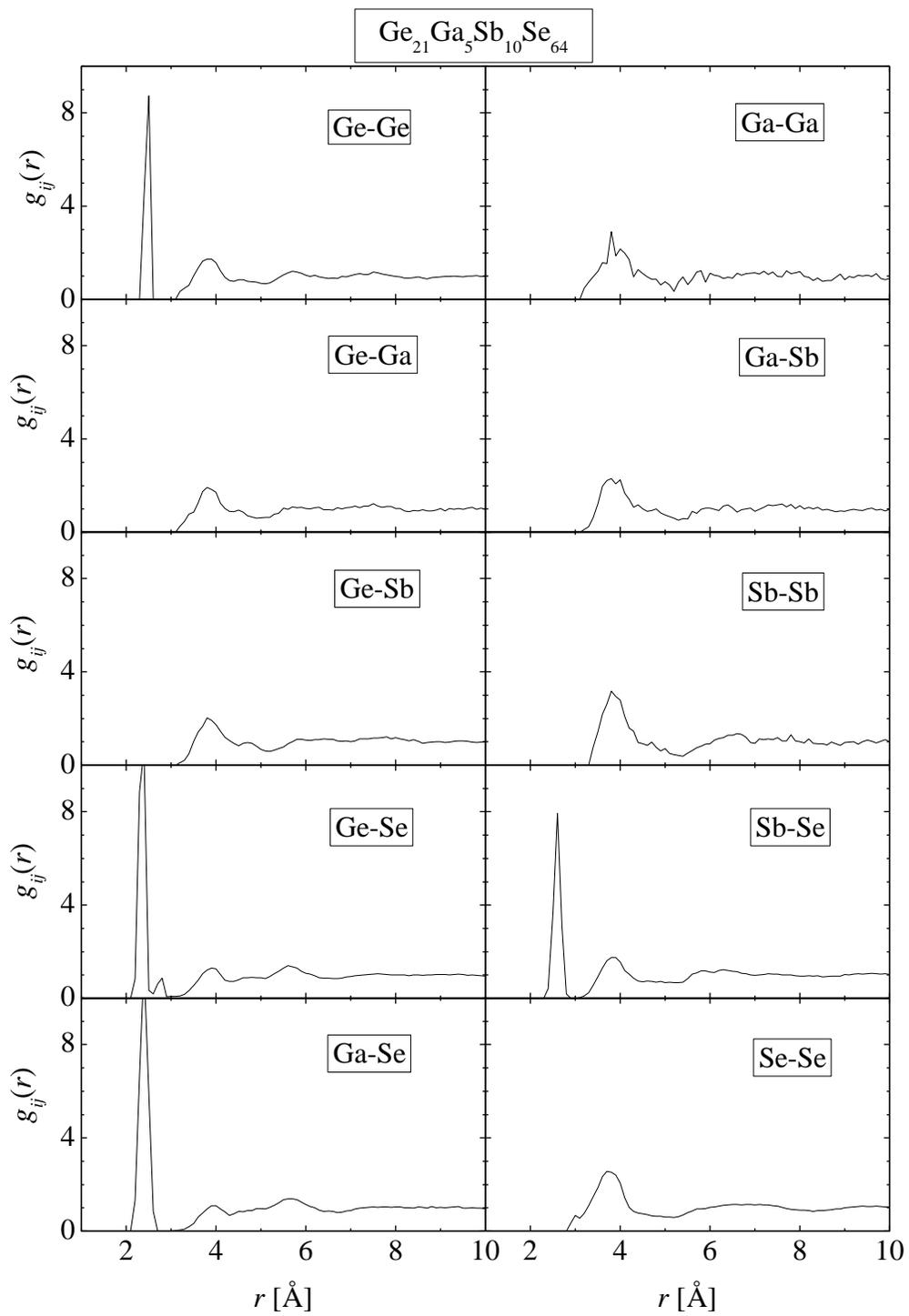

Figure 3 Partial pair correlation functions of the $Ge_{21}Ga_5Sb_{10}Se_{64}$ sample.

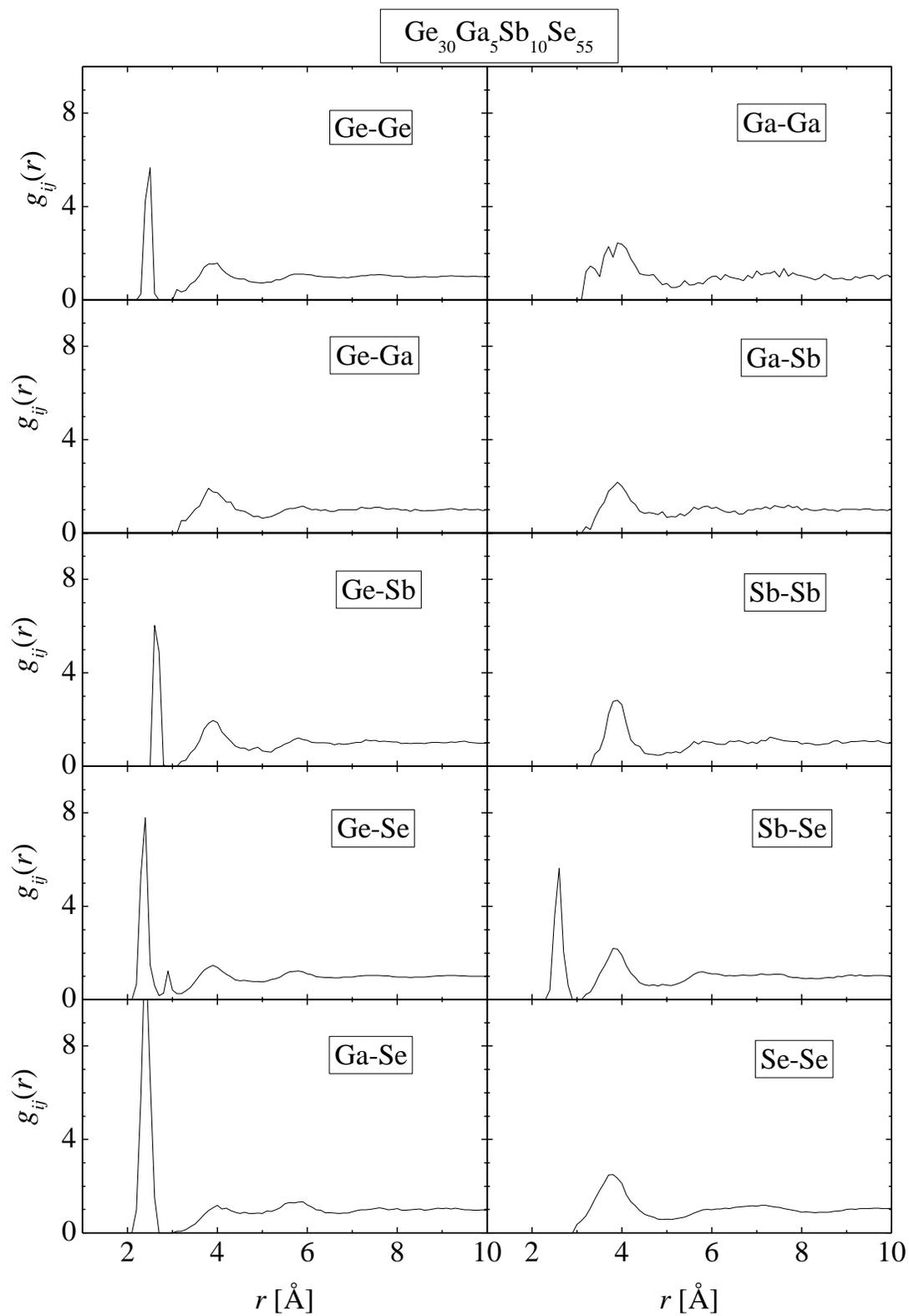

Figure 4 Partial pair correlation functions of the $Ge_{30}Ga_5Sb_{10}Se_{55}$ sample.

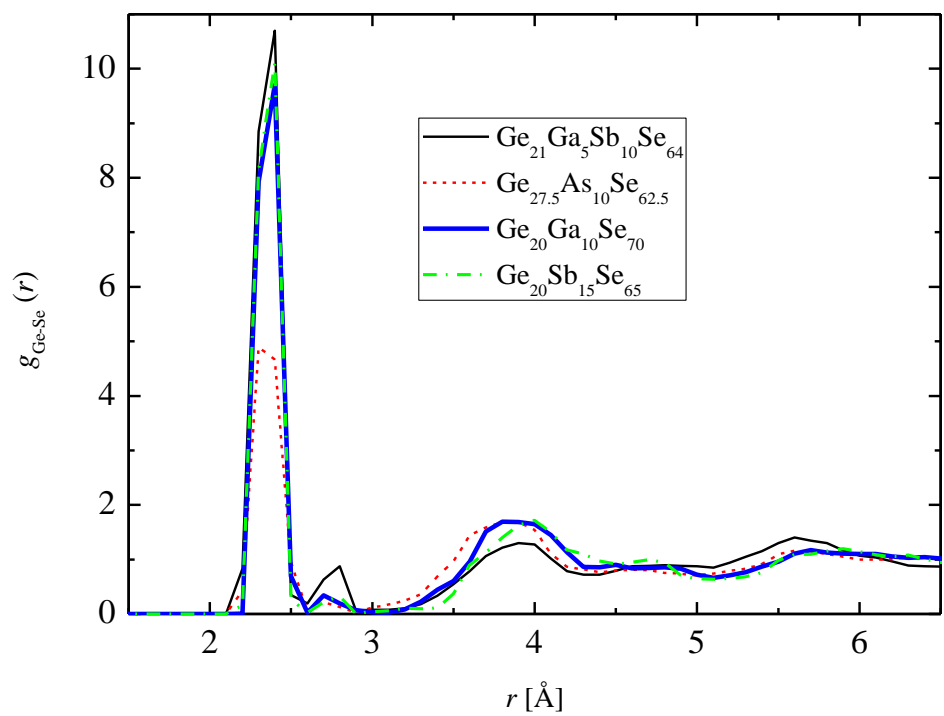

Figure 5 Comparison of Ge-Se partial pair correlation functions of $Ge_{21}Ga_5Sb_{10}Se_{64}$, $Ge_{27.5}As_{10}Se_{67.5}$, $Ge_{20}Sb_{15}Se_{65}$ and $Ge_{20}Ga_{10}Se_{70}$.